\documentclass[aps,twocolumn,prb,floatfix,groupedaddress,showpacs,longbibliography]{revtex4-1}
\usepackage[utf8]{inputenc} 
\usepackage{amssymb}
\usepackage{epsf}
\usepackage{epsfig}
\usepackage{graphicx}
\usepackage{dcolumn}
\usepackage{braket}
\usepackage{bm}
\usepackage{amsfonts}
\usepackage{amsmath}
\usepackage{amssymb}
\usepackage{color,soul}
\usepackage{wasysym}
\usepackage{mathrsfs}
\usepackage{times}
\usepackage[dvipsnames,svgnames,table]{xcolor}
\usepackage{ulem}
\usepackage{hyperref}
\usepackage{fancyhdr}
\usepackage{float}
\usepackage[english]{babel}
\usepackage{hyperref}

\hypersetup{
    unicode=true,          
    pdftoolbar=true,        
    pdfmenubar=true,        
    pdffitwindow=false,     
    pdfstartview={FitH},    
    pdftitle={},    
    pdfauthor={},     	
    colorlinks=true,       	
    linkcolor=NavyBlue,  
    citecolor=Maroon,       
    filecolor=NavyBlue,      
    urlcolor=NavyBlue       
}

\newcommand{\bea}{\begin{eqnarray}}
\newcommand{\eea}{\end{eqnarray}}

\begin{document}

\title{Multi-terminal Conductance at the Surface of a Weyl Semimetal}
\author{J. Chesta Lopez}
\affiliation{Departamento de F\'{\i}sica, Facultad de Ciencias F\'{\i}sicas y Matem\'aticas, Universidad de Chile, Santiago, Chile}
\author{L. E. F. Foa Torres}
\affiliation{Departamento de F\'{\i}sica, Facultad de Ciencias F\'{\i}sicas y Matem\'aticas, Universidad de Chile, Santiago, Chile}
\author{A. S. Nunez}
\affiliation{Departamento de F\'{\i}sica, Facultad de Ciencias F\'{\i}sicas y Matem\'aticas, Universidad de Chile, Santiago, Chile}

\begin{abstract}
Weyl semimetals are a new paradigmatic topological phase of matter featuring a gapless spectrum. One of its most distinctive features is the presence of Fermi arc surface states.  
Here, we report on atomistic simulations of the dc conductance and quantum Hall response of a minimal Weyl semimetal. By using scattering theory we show that a quantized Hall conductance with a non-vanishing longitudinal conductance emerges associated to the Fermi arc surface states with a remarkable robustness to high concentrations of defects in the system.
Additionally, we predict that a slab of a Weyl semimetal with broken time-reversal symmetry bears persistent currents fully determined by the system size and the lattice parameters.

\end{abstract}
\date{\today}
\maketitle

\section{Introduction} 

Topological but gapless, Weyl semimetals~\cite{hosur_recent_2013,vafek_dirac_2014} challenge our physical intuition with striking properties, thereby constituting a new paradigm in the physics of topological quantum matter. 
One of their distinctive features is the presence of Weyl nodes, discrete points in the three-dimensional (3D) Brillouin zone where the valence and conduction bands meet forming a 3D analog of graphene. Unlike the Dirac points in graphene, Weyl nodes are non-degenerate and a perturbation does not gap the system, it only shifts the Weyl nodes in \textbf{k}-space~\cite{YanFelser2017}. Interestingly, the existence of Weyl nodes in the sample's bulk dispersion, which appears in pairs of opposite chirality acting as Berry phase monopoles, can be tied to the existence of Fermi arc surface states at the boundary of a finite system. Control over the exotic electronic properties of a Weyl semimetal is expected to lead to the development of novel devices. In particular, high temperature quantum devices are envisioned to take advantage of the topological protection of their states. Superconducting Weyl semimetals, expected to host novel forms of Majorana fermions, are likely to aid in the development of quantum information devices.~\cite{jia-commentary,chiral-anom-use,photoinduced-use,floquet-use,photogalvanic-use}

Fueled by the experimental observation of Weyl semimetals in recent years~\cite{xu_discovery_2015,lu_experimental_2015,lv_experimental_2015}, a plethora of different flavors is now known to exist~\cite{bradlyn_beyond_2016} and there is a fierce competition to unveil them in the laboratory. While most experiments rely heavily on probing the electronic dispersion through angle resolved photoemission spectroscopy (ARPES), transport experiments are also expected to provide for valuable information~\cite{reis_search_2016,belopolski_discovery_2016}. In the bulk, the conductivity tensor is expected to be anisotropic and, depending on the crystal symmetry, a giant anomalous Hall effect may arise~\cite{yang_quantum_2011} ($\sigma_{yz}=\frac{e^2}{2\pi h} (k_+-k_-)$, with $k_+-k_-$ being the momentum separation between the Weyl nodes).

In spite of the intense research, with few exceptions~\cite{baireuther_scattering_2016,igarashi_magnetotransport_2017,3DquantumHall}, most studies have focused on the bulk rather than the surface transport properties of Weyl semimetals~\cite{yang_quantum_2011,hosur_charge_2012,Nandy2017}. Besides being an easier to access property than its bulk counterpart, the transport response measured at the surface may guard new hallmarks of the Weyl semimetal phase. Theoretical simulations may help to unveil them while clarifying many interesting aspects that are unique to this topological phase. For example, in contrast to topological insulators, the bulk states in Weyl Semimetals are expected to play a role in transport~\cite{gorbar_origin_2016}. But the intricate electronic structure of many Weyl semimetals featuring sometimes dozens of Weyl nodes poses an obstacle in this young field. Studying a simple ''hydrogen-like'' model of a Weyl semimetal is, therefore, a first necessary step.

Here, we report on the electronic and transport properties of a minimal Weyl semimetal with only two Weyl nodes. This minimal model, which requires broken time-reversal symmetry, is used as a playground for the study of the multiterminal conductance at the surface of a finite slab within a coherent approach. To such end, we combine a scattering formulation of transport together with a tight-binding model. Our simulations show that associated with the Fermi arc surface states, there is a nearly quantized Hall conductance which remains robust to imperfections and defects. Furthermore, and unlike to Chern insulators, this Hall response is accompanied by a non-vanishing longitudinal conductance. These two observations, the deviations of the Hall conductance from perfect quantization and the non-vanishing longitudinal conductance, are a consequence of the unavoidable leak of the surface states into the bulk. We also test the topological nature of our quantized conductance results by adding random lattice defects and find that they remain robust at concentrations of the order of 25\% which is remarkably higher than what has been previously reported on 2D topological systems~\cite{Castro2015}.

Interestingly, our results also show that a slab of a Weyl semimetal can bear \textit{persistent currents} even in the absence of a magnetic field. These currents are governed by the system size and the intrinsic time-reversal symmetry breaking lattice parameters and can explain why the surface transport is impeded in the direction perpendicular to the Fermi arcs.

\section{Hamiltonian model}

For our calculations we use a model of spinful electrons on a cubic lattice (one orbital per site with lattice constant $a$)~\cite{yang_quantum_2011,vafek_dirac_2014},
\begin{equation}
\label{Hamiltonian}
\begin{aligned}
{\cal H}_k &= [ 2t_x \left( \cos(k_x) - \cos(k_0) \right) +\\
& + m \left(2-\cos(k_y)-\cos(k_z)\right)]\sigma_x + \\
& +2t_y \sin(k_y)\sigma_y + 2 t_z \sin(k_z)\sigma_z.
\end{aligned}
\end{equation}

\begin{figure}[h]
\includegraphics[width=0.45\textwidth]{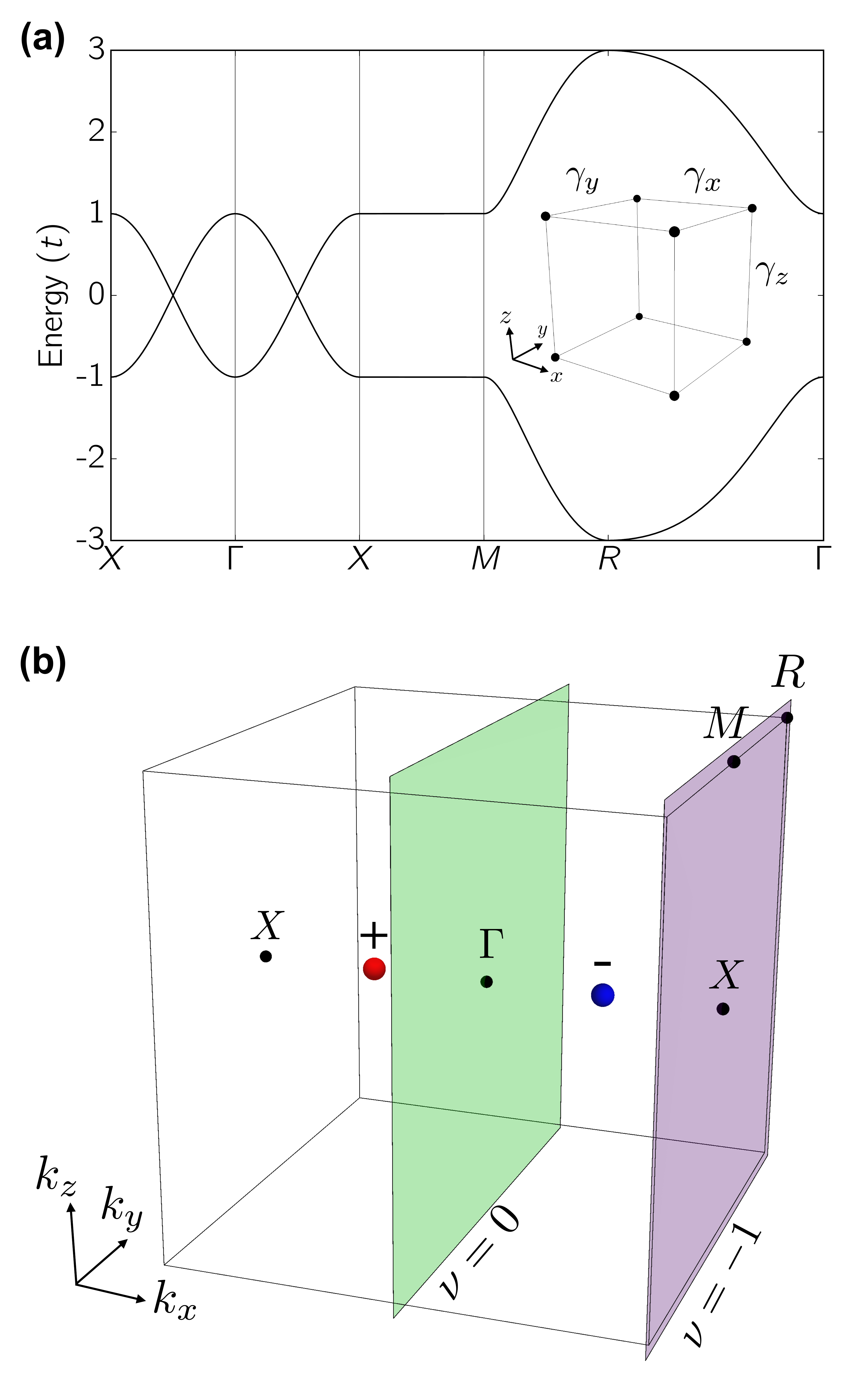}
\caption{(a) Bulk dispersion of the semimetal in the topological phase with only two Weyl nodes with parameters $m=1.0$, and $t_x=t_y=t_z=1/2$. The inset shows a cell of the real space lattice with nearest neighbor hoppings $\gamma_x$, $\gamma_y$, $\gamma_z$. (b) First Brillouin zone of the Weyl semimetal showing the high symmetry points where the dispersion in (a) was calculated. We show also the Weyl nodes of opposite chirality, and the planes where the Chern numbers were obtained.}
\label{fig1}
\end{figure}

As noticed in Ref.~\onlinecite{yang_quantum_2011}, this model has two, four, six or eight Weyl nodes depending on the interplay between the hoppings $t_x$, $t_y$, $t_z$ and the mass term determined by $m$.
Each pair of nodes is related by \textit{inversion symmetry} which is preserved as there is a single orbital per site and only one site in the lattice basis.

\begin{figure}[htp]
\includegraphics[width=0.5\textwidth]{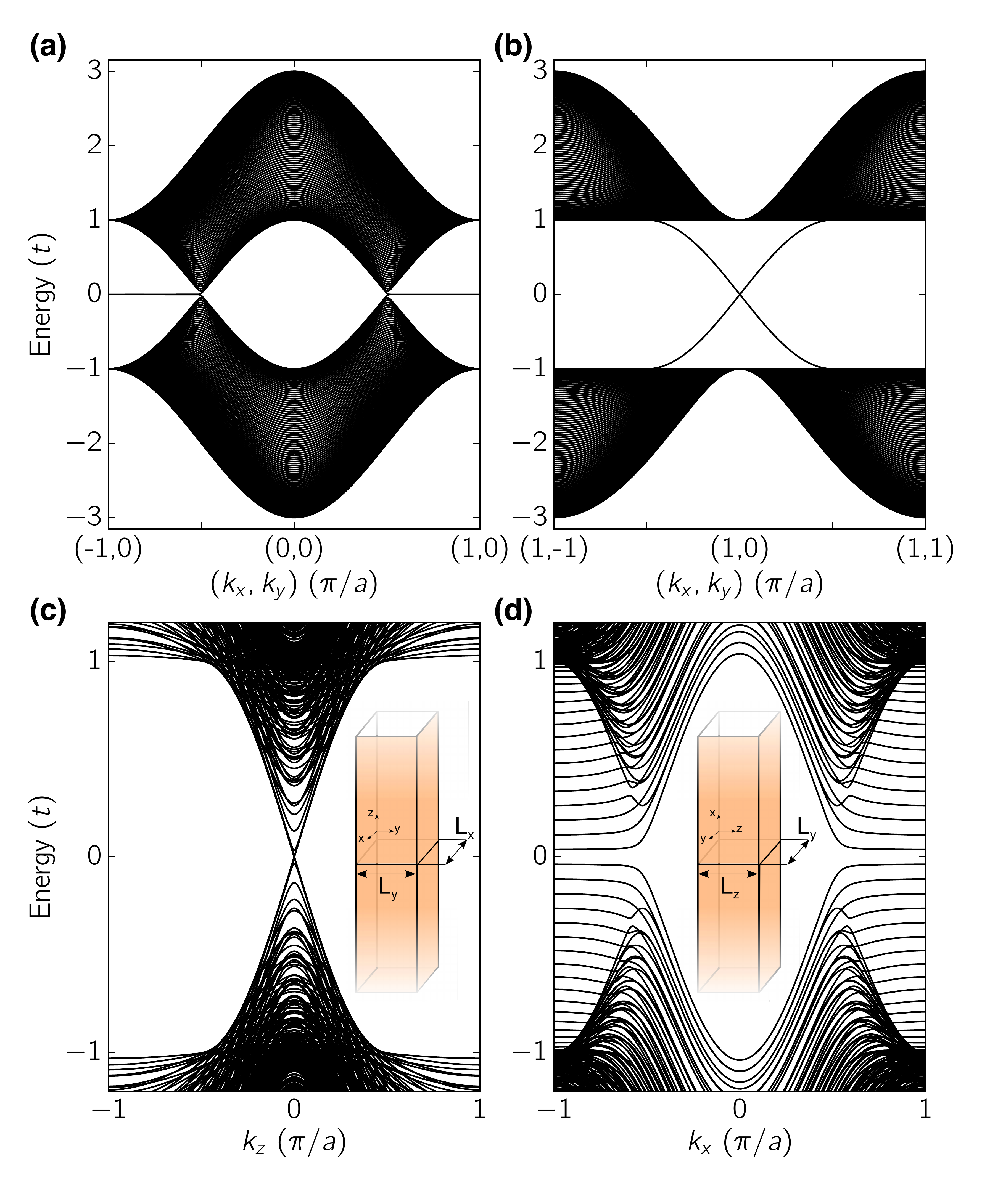}
\caption{Energy dispersions for the model discussed in the text. (a) and (b) correspond to a system with 100 lattice sites in the z-direction (and infinite in $x$ and $y$). (a) shows the dispersion along the $k_y=0$ line in the 2D Brillouin zone, while (b) shows it along the $k_x=\pi/a$ line. (c) and (d) show the energy dispersion for a Weyl semimetal wire with a cross section of $30 \times 30$ lattice sites (with the same parameters as in (a) and (b)). (c) corresponds to a wire with the cross-section in the $x-y$ plane (an $x-z$ wire has identical dispersion) while (d) corresponds to a wire with the cross-section in the $y-z$ plane. The insets show the geometry of the wire in each case.}
\label{fig2}
\end{figure}

The breaking of the \textit{time reversal symmetry} is governed by the parameter $m$ in the Hamiltonian. Let us analyze this in more detail. The time-reversal operator $\cal{T}$ can be written as $\cal{T}=U\theta$, where $\cal{U}$ is a unitary operator and $\theta$ is the complex conjugation operator ($\theta^{-1}=\theta$ (see for example~\cite{haake_time_2001}, chapter 2), one can also show that $\theta \sigma_x \theta=\sigma_x$, $\theta \sigma_y \theta=-\sigma_y$ and $\theta \sigma_z \theta=\sigma_z$). For spin $1/2$ particles, the unitary operator must have the form: ${\cal U}=e^{i\phi}\sigma_y$, with $\phi$ an arbitrary phase. This satisfies all the required conditions for a time-reversal operator: it is anti-unitary, changes the sign of spin and is a $2\times 2$ matrix. The second condition can be expressed as: ${\cal T} \sigma_j {\cal T}^{\dagger}=-\sigma_j$.

\
\\
\section{Electronic Structure}

For the electronic structure and transport calculations that follow we set $t_x = t_y = t_z = 1/2$ and $m=1.0$ so that our model contains only two Weyl nodes located at $\vec{k}=(\pm \pi/2a,0,0)$ as shown in the bulk dispersion calculated along high symmetry points in Fig.~\ref{fig1}(a). This phase is characterized by a Chern number $\nu=0$ in the 2D Brillouin planes located between the Weyl nodes, and a Chern number $\nu=-1$ in the planes located outside of the nodes (see Fig.~\ref{fig1}(b)). The topological nature of this phase motivates us to ask about the transport response originated by the surface states in this minimal case, but before turning to this question, we need to review its electronic structure in more detail.

\begin{figure*}[t]
\includegraphics[width=0.95\textwidth]{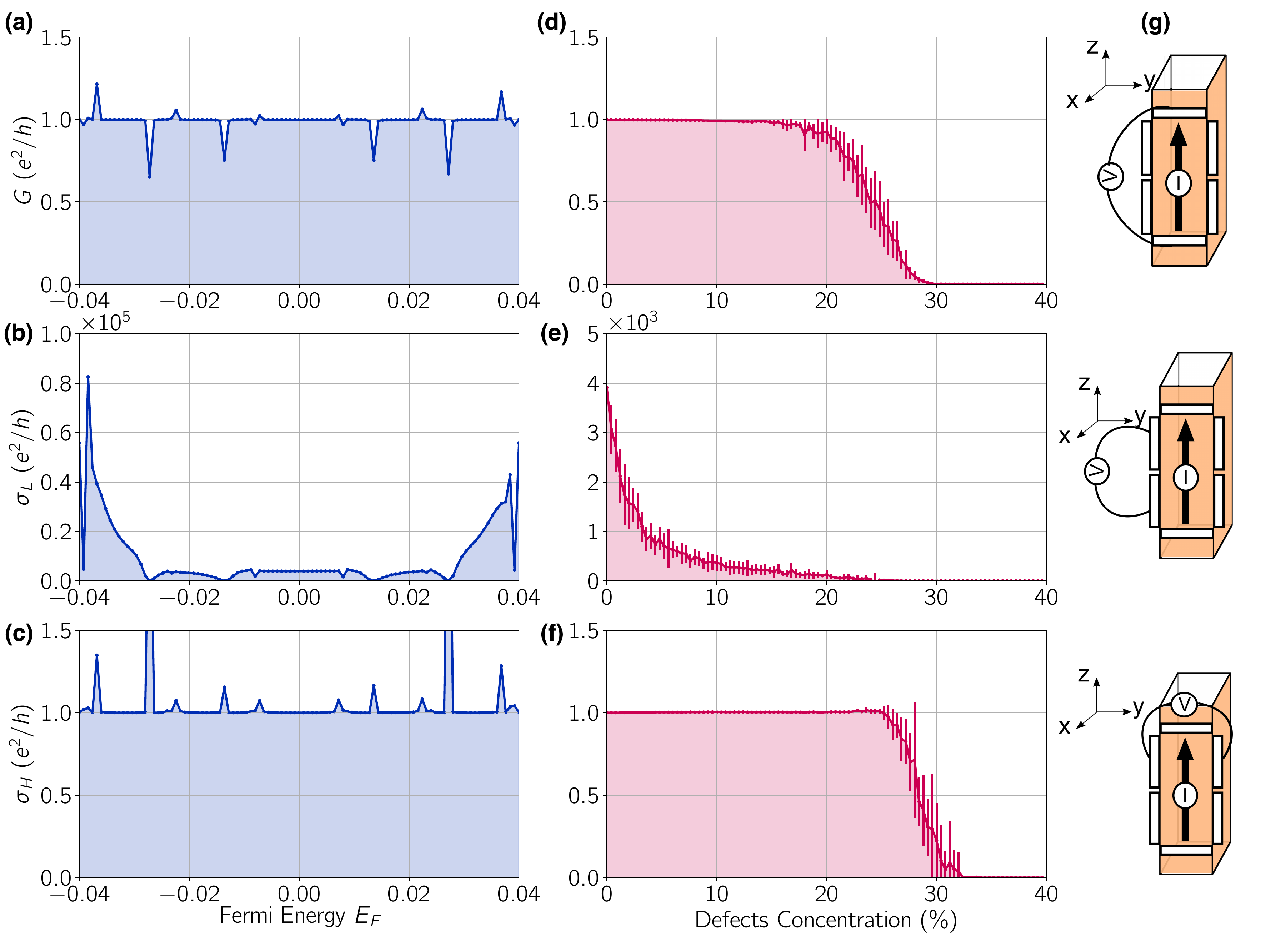}
\caption{Transport response at the surface of the $yz$ face of a rectangular cuboid. Six electrical contacts in an H configuration access the yz plane as shown in (g). (a) Two-terminal conductance $G$, (b) Longitudinal conductance $\sigma_{L}$, and (c) Hall conductance $\sigma_{H}$, are plotted against the Fermi Energy $E_f$ set on the leads. Panels (d), (e) and (f) show the same magnitudes as their counterparts on the left but averaged over $15$ indepedent realizations of disorder as a function of the concentration of defects (vacancies) and for $E_F=0.001$. The bars depict the fluctuations around the average values. The setups are represented in the schemes in (g), note that the contacts are two-dimensional.}
\label{fig3}
\end{figure*}

The numerical calculations were implemented in Python using the modules: Kwant~\cite{groth_kwant:_2014} and Python tight-binding
~\footnote{The Python Tight-Binding package is available at \href{http://physics.rutgers.edu/pythtb/}{this link}}.

As mentioned earlier, the Fermi arcs are gapless bands that connect the projections of the Weyl nodes in a given plane of the Brillouin zone. Since we are in a phase with a single pair of nodes, only one Fermi arc is to be expected on this surface. To evidence the Fermi arcs we need to terminate the system at the correct surface. Figure~\ref{fig2}(a) and (b) show the band structure for a system with $100$ lattice sites in the z-direction (and infinite on the other two). In Fig.~\ref{fig2}(a), the two gapless bands at zero energy form the Fermi arc. Nevertheless, these dispersionless bands associated to the Fermi arc in the $k_x$ direction do have a non-vanishing group velocity along $k_y$ as shown in Fig.~\ref{fig2}(b).

Next, we focus on the electronic structure of a Weyl semimetal wire with a square cross-section of $30\times 30$ lattice sites in the $xy$, $xz$ and $yz$ planes respectively. 
The band structures for the wires are shown in Fig.~\ref{fig2}(c) and (d). Figure \ref{fig2}(c) corresponds to a wire extending along $z$ (i.e. with the cross-section in the $xy$ plane), the results for a wire extending along $y$ are identical. The projection of the two Weyl Nodes is a single Dirac cone that is doubly degenerate and hosts edge states around its center. The wire with the cross-section in the yz plane has a different band structure as shown in Fig.~\ref{fig2}(d). Flat bands appear around $E_F=0$ that do not carry associated edge states. These flat bands are the manifestation of the Fermi arcs: due to the to finite size effects of the wire geometry they present a splitting which scales with $1/N^2$.

\
\\

\section{Multiterminal transport response} 

\begin{figure}[h]
\includegraphics[width=0.5\textwidth]{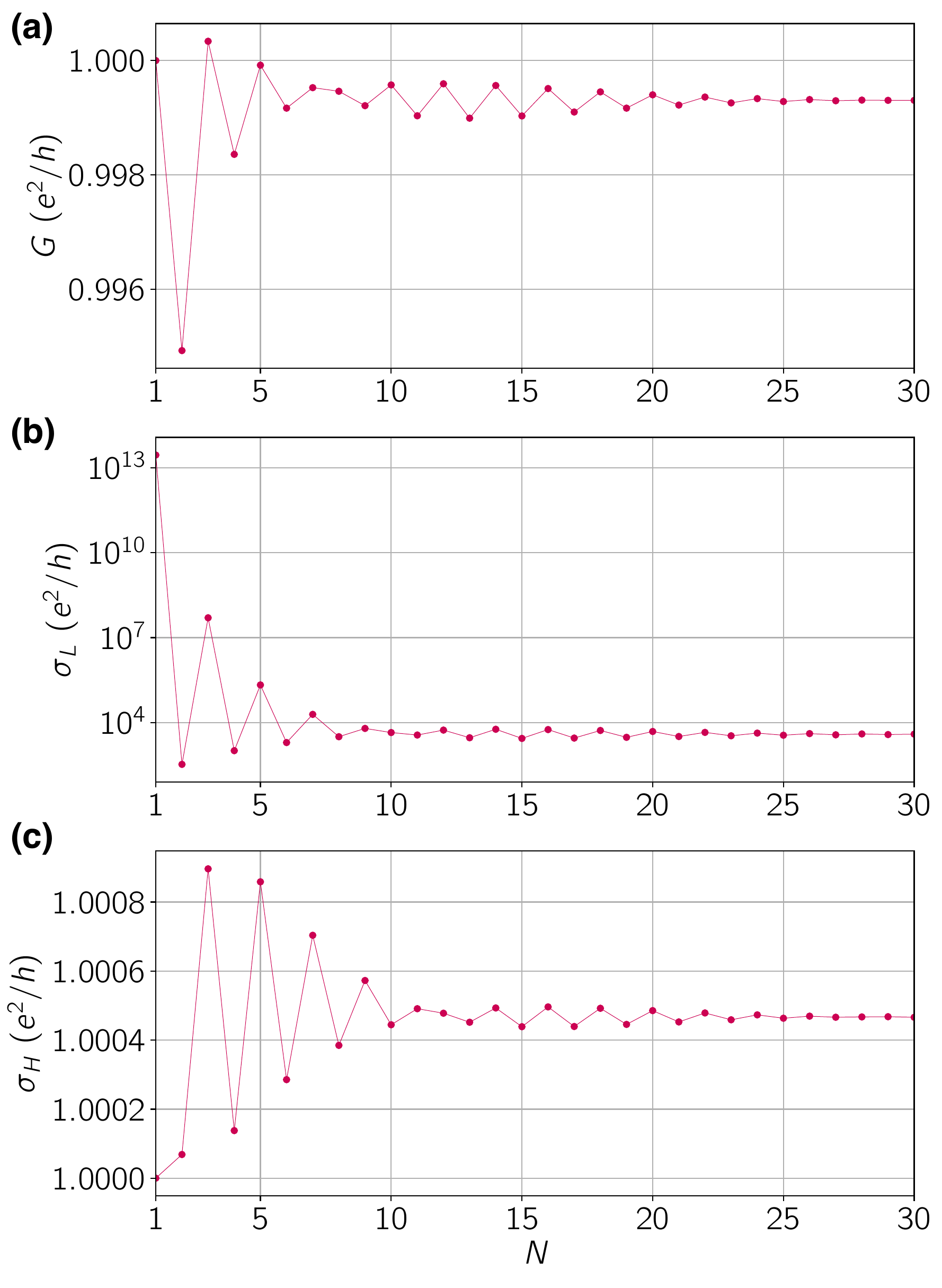}
\caption{In this figure we explore how $G$ (a), $\sigma_{L}$ (b), and $\sigma_{H}$ (c), determined from probes on the $yz$ face as in the previous figure, change as the dimension of the Weyl semimetal along $x$, $N$, is varied. Here, we set $E_F=0.001$.}
\label{fig4}
\end{figure}

Let us now turn to the transport response of this `hydrogen-model' of a Weyl semimetal. We are interested, in particular, in a Hall setup at one of the surfaces of a three-dimensional sample. Specifically, we consider six two-dimensional electrodes connected (in an H-configuration) to the surface of a rectangular cuboid as represented in Fig.~\ref{fig3}(g). Note that this differs from other works where the sample is accessed with the electrodes on the full volume~\cite{Takane2016,Kobayashi2015} and not just the surface. In the following we present simulations for a rectangular cuboid of $180 \times 30 \times 30$ lattice sites in the $x$, $y$ and $z$-directions, respectively. Each lead is described by a two-dimensional square lattice with lattice constant matching that of the 3D cuboid and $30$ lattice sites in width. The separation between the voltage probes is set to $120$ lattice sites. The 2D lattice has nearest neighbors and spin independent hoppings equal to $t_{lead}=1.0$. This value also characterizes the hopping from sites in the leads to the matching points in our structure. 

The setup studied here (see scheme in Fig.~\ref{fig3}(g)) is typically used in Hall measurements and allow for the determination of the longitudinal and Hall resistances associated with the edge states. 
Furthermore, a planar architecture for the contacts may offer advantages, such as easier gating and tuning, over three-dimensional counterparts. Such contacts problems together with difficulty in tuning the Weyl semimetals themselves~\cite{belopolski_discovery_2016}.
In our case, we would like to determine the fingerprints of the topology of this 3D system in the transport response measured at a surface. The answer is a priori not obvious since the states associated to the Fermi arcs are surface states, and not edge states as in a Chern insulator. Interestingly, one could also study the crossover between those two cases by varying the dimension of the cuboid so that it becomes 2D (we will come back to this when discussing Fig.~\ref{fig4}), a transition that has elicited much interest~\cite{Burkov2011,Kazuki2017}.

Our numerical simulations are restricted to the coherent regime, where scattering theory is valid. Specifically, we apply the Landauer-Büttiker formalism~\cite{datta_electronic_1995} to this three-dimensional system (see schemes in Fig.~\ref{fig3}(g)) with the aid of the Kwant module~\cite{groth_kwant:_2014}. Fig.~\ref{fig3} shows the two-terminal (panel a), longitudinal (panel b) and Hall conductances (panel c) as a function of the Fermi energy in the zero-temperature limit. These results correspond to a setup where the leads are attached to the $yz$ surface. We emphasize that our results (as well as those of Ref.~\cite{Takane2016}) correspond to conductances rather than conductivities as calculated by other authors~\cite{Chen2015,Shapourian2016}. In the latter case, since they are bulk quantities, the effect of Fermi arcs are less evident.

\begin{figure}[h]
\includegraphics[width=0.5\textwidth]{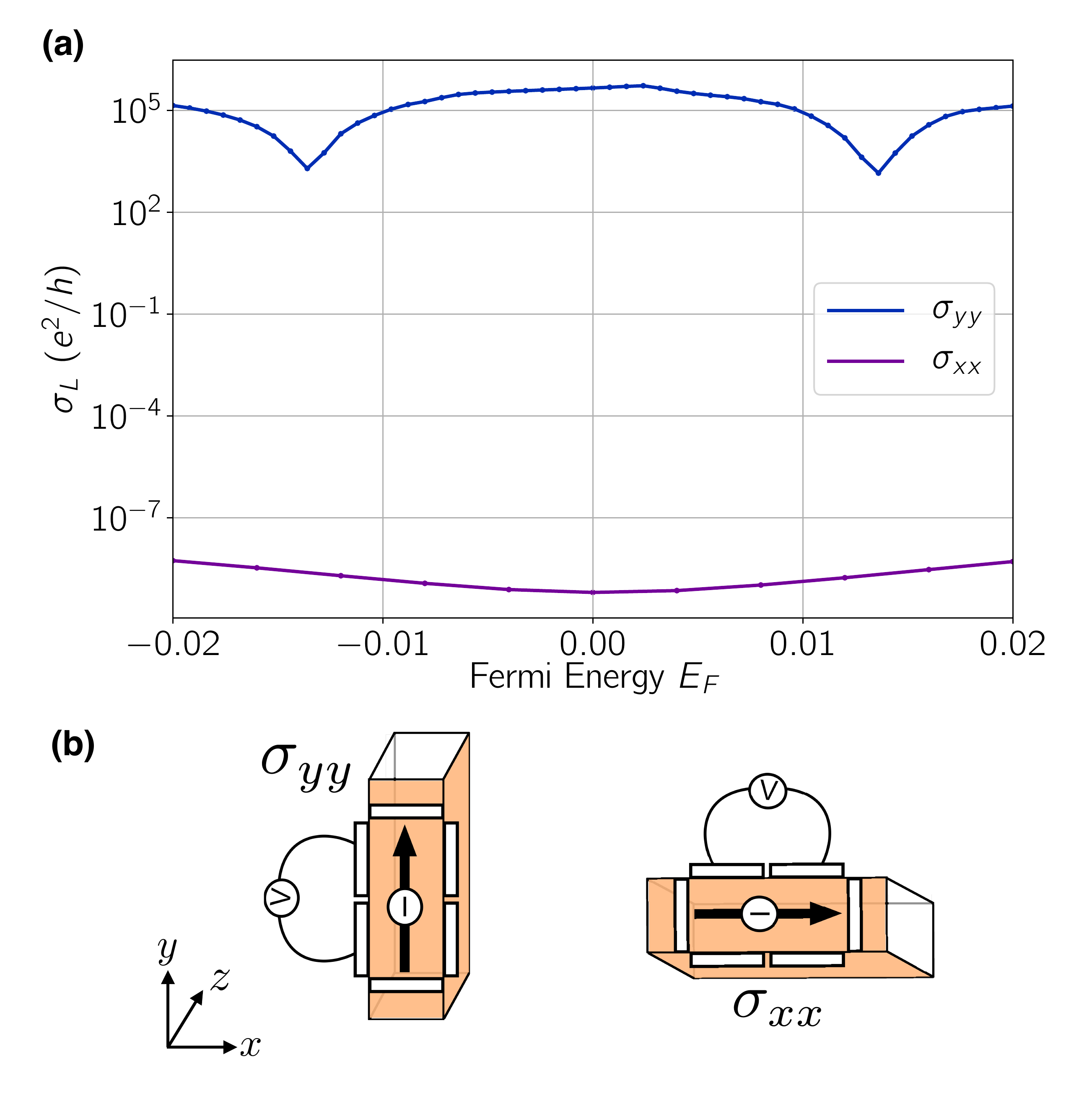}
\caption{Transport response at the surface of the xy face of a rectangular cuboid. Six electrical contacts in an H configuration access the xy plane. (a) Longitudinal conductance $\sigma_{L}$ is plotted against the Fermi Energy $E_F$ set on the leads for two different orientations of the cuboid geometry, thus measuring either $\sigma_{yy}$ or $\sigma_{xx}$ as shown in panel (b).}
\label{fig5}
\end{figure}

In the first place, as shown in Fig.~\ref{fig3}(c), we find a nearly-quantized Hall conductance $\sigma_{yz}  \simeq \frac{e^2}{h}$. This value of the Hall conductance is connected with the Chern number $\nu=-1$ on the plane in the Brillouin zone that is outside the Weyl nodes because the planes with non-zero Chern number can be regarded as the stacking of 2D Chern insulators in a quantum Hall effect phase~\cite{yang_quantum_2011}. The differences between the simulated values and perfect quantization are in the order of $10^{-7}$, this is $\sigma_{yz}-\frac{e^2}{h} \simeq \times 10^{-7}$, except at a few discrete points that coincide with the onset of new subbands either in the semimetal or in the leads.

\begin{figure}[h] 
\includegraphics[width=0.5\textwidth]{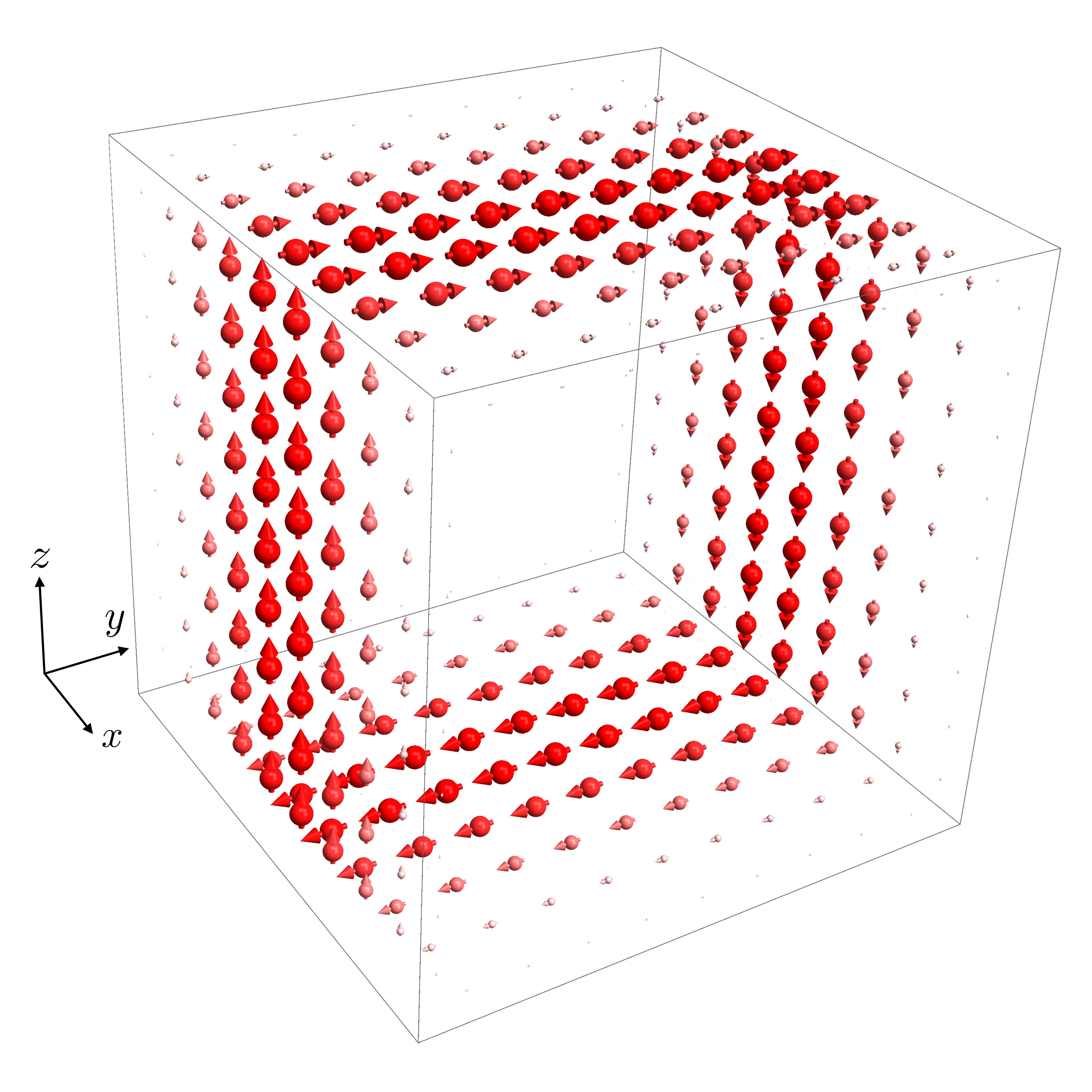}
\caption{Scheme showing the current density flow in a finite sample of Weyl Semimetal $10\times 10 \times 10$ lattice sites, at $E_F=0$. The arrows point in the direction of the current flow, and their size and color intensity are proportional to the magnitude of the current density.}
\label{fig6}
\end{figure}
The non-vanishing value of the Hall conductance even in the absence of an external magnetic field is a consequence of the intrinsic breaking of time reversal symmetry in our model Hamiltonian, i.e. by the role played by the parameter $m$, much as in the (anomalous) quantum Hall effect~\cite{haldane_model_1988,Nagaosa2010}. But besides this similarity, there are also differences that we point out below. They concern the longitudinal conductivity and the robustness to disorder.

In addition, we observe a non-vanishing longitudinal conductance as shown in Fig.~\ref{fig3}(b). This is connected with the observed small deviations from perfect quantization of the Hall conductance and can be interpreted in terms of the finite and non-vanishing penetration length of the surface states, which decay exponentially into the bulk.

To get further insight, we test the robustness of these results to defects. To such end, we add vacancies at random within to the sample and computed how the two-terminal, longitudinal and Hall conductances changed in the same setup. The results are shown in the panels (d), (e) and (f) of Fig.~\ref{fig3}, averaged $15$ realizations of disorder. For a given defects concentration one chooses a set of sites and transform them into vacancies. The results turn out to be remarkably robust to disorder, with the conductance being substantially reduced only at about 25\% of defects. This is much higher than expected, for example, in the two-dimensional Haldane model~\cite{Castro2015}. 

Indeed, an interesting question is how does the transport response evolve as the cuboid becomes thinner and, eventually, two-dimensional. In the present setup this can be done without changing the measurement probes which are on the surface. Fig.~\ref{fig4} shows the results for the two-terminal conductance $G$, the longitudinal conductance $\sigma_L$, and the hall conductance $\sigma_H$ (computed at $E_F=0.001$) as a function of the dimension of the cuboid along $x$, $N$. $N$ spans from the two-dimensional limit ($N=1$) to 
$N=30$ as used in Fig.~\ref{fig3}.

Figure~\ref{fig4} shows that in the two-dimensional limit transport occurs as in a Chern insulator, with the Hall conductance being perfectly quantized and a vanishing longitudinal resistance. This picture changes as the sample becomes thicker until it converges to a situation where the Hall conductance aparts itself slightly from the perfect quantization with the concomitant non-vanishing longitudinal resistance. Besides clarifying the effect of the sample size on our previous results, this figure highlights the role of the dimensionality and the effect of the penetration of the surface states into the bulk.

To complete our analysis we study the transport response on the $xy$ plane (the same results are obtained for the $xz$ plane). We find that there is no Hall conductance response and that $\sigma_{xx}\ll\sigma_{yy}$ as shown in Fig.~\ref{fig5}(a). These results are a consequence of the dispersion shown in Fig.~\ref{fig2}(a) and (b). The Fermi arcs have group velocity in the $y$ and $z$ directions only, and thus do not allow for transport in the $x$ direction. The previous results motivate us to analyze how do the currents flow in a finite sample of the semimetal. Since transport is impeded in the $x$ direction, there also should not be current density flowing along $x$.

\begin{figure}[h]
\includegraphics[width=0.5\textwidth]{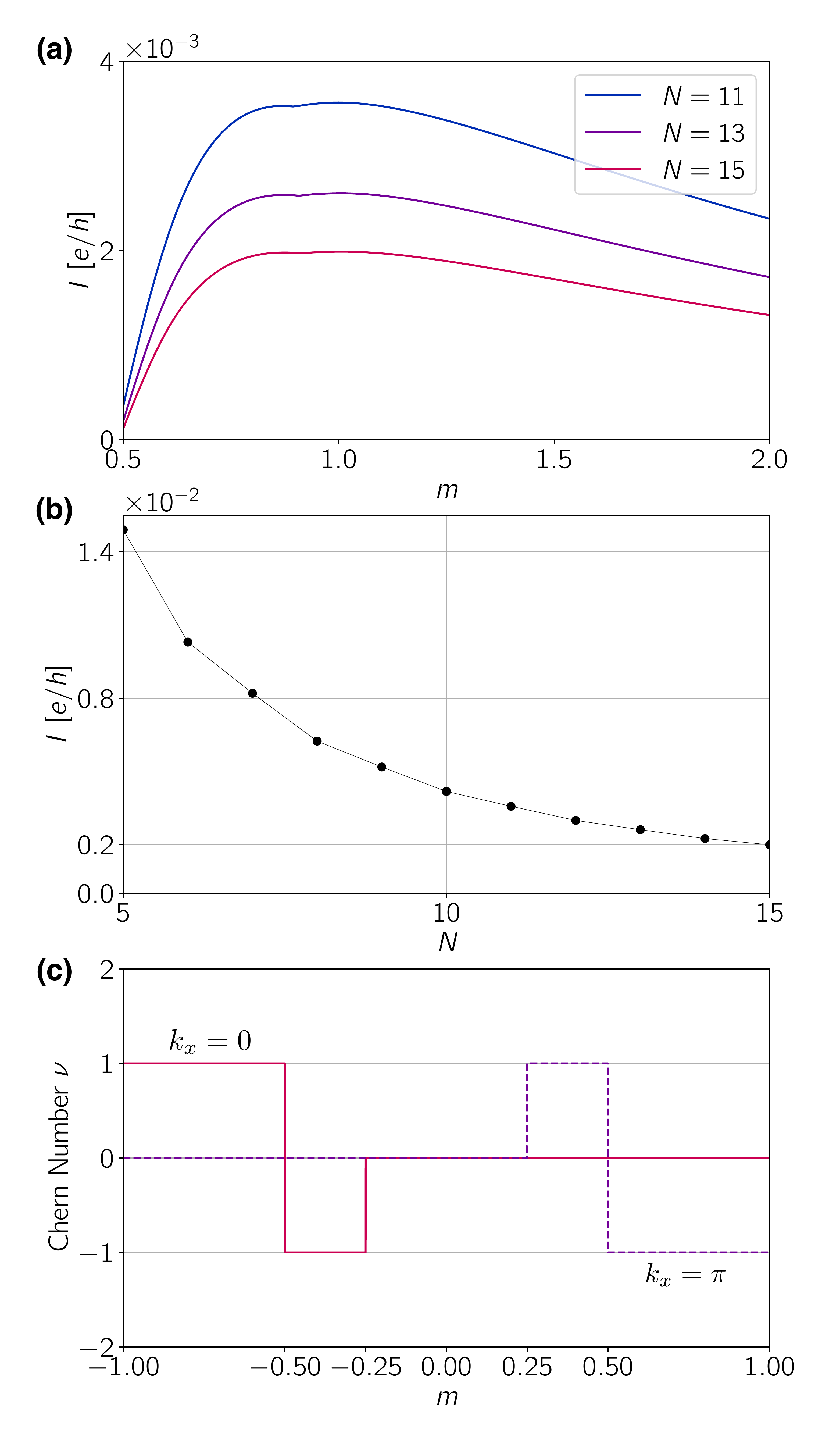}
\caption{(a) Maximum surface current for a cube of $N\times N\times N$ lattice sites as a function of $m$, the time reversal breaking parameter. (b) Maximum surface current as a function of the cube lateral dimension $N$ at fixed $m=1.0$. (c) Values of the Chern number $\nu$ for the planes intersecting $k_x=0$ and $k_x=\pi/a$ in the Brillouin zone, as a function of $m$. }
\label{fig7}
\end{figure}

\
\\
\\
\section{Chiral States and Persistent Currents}
Inspired by the previous conclusion, we calculate the bond currents~\cite{todorov_tight-binding_2002} at $E_f=0$ for a finite sample of the semimetal. We find that even in the absence of an external magnetic field, the system develops persistent currents around the $x$ axis. These currents flow through the boundary of the semimetal as shown schematically in Fig.~\ref{fig6}. We attribute these results to the time-reversal breaking acting as an intrinsic magnetic field pointing parallel to the Fermi arcs. The currents in the system are only of the solenoid type; transport along the $x$ axis is impeded in the pristine system (the states at the boundary are dispersionless), thereby producing a very high longitudinal resistance in the $x$ direction compared to the other directions. This picture helps to explain our transport results.

The above picture is still not quantative. Figure \ref{fig7} (a) shows the magnitude of the maximum surface current (maximum bond current) as a function of the time reversal breaking parameter $m$ (Fig. \ref{fig7} (c) shows the Chern numbers for two different planes in the Brillouin zone as a function of $m$). Different curves correspond to different system sizes of a $N\times N\times N$ cube. The horizontal axis ranged at $m=0.5$, the value which signals the beginning of the topological phase. Figure \ref{fig7} (b) shows a detail of the current dependence with the system size at fixed $m=1.0$.

\
\\
\\
\section{Final Remarks} 
In summary, we have explored the multiterminal transport properties of a minimal model of a Weyl semimetal with broken time-reversal symmetry. This kind of hydrogen-atom model of a Weyl semimetal allows for a simplified analysis. Our results show a nearly quantized Hall conductance, a result that holds up to remarkably high concentrations of disorder. Interestingly, in contrast with the quantum Hall effect studied in time-reversal invariant Weyl semimetals, the Hall response in our  time-reversal broken model is not associated to formation of edge states (at the surface of the Weyl semimetal) but rather to the surface states.

Being impervious to disorder, the transport properties of the surfaces of a Weyl semimetal have a great potential in applications in nanotechnological devices. In this way, our calculations provide the motivation and are the starting point of new simulations required to elucidate the role of finite geometries upon the transport properties of more complex Weyl semimetals.

\vspace{1cm}

\vspace{0.25cm}
\noindent
\textit{Acknowledgments.--} We acknowledge financial support from Universidad de Chile through Program 'Inserci\'on Acad\'emica 2016. ASN acknowledges funding from grants Fondecyt Regular 1150072 and from Financiamiento Basal para Centros Científicos y Tecnológicos de Excelencia, under Project No. FB 0807, CEDENNA (Chile). LEFFT acknowledges the support of Fondecyt Regular 1170917.\\
\end{document}